\documentclass{PoS}

\title{Diffractive asymmetry of electroweak vector bosons at the LHC}

\ShortTitle{Diffractive hadroproduction of vector bosons}

\author{\speaker{Agnieszka \L{}uszczak}\thanks{This work is partially supported by the grant of MNiSW 
N202 249235.}\\
        Institute of Nuclear Physics Polish Academy of Sciences, Cracow, Poland\\
        E-mail: \email{agnieszka.luszczak@ifj.edu.pl}}

\abstract{We analyse diffractive weak boson production in $pp$ collisions. We show that the measurement of $W$ boson production asymmetry in the diffractive $pp$ collisions is a valuable method to test the concept of the flavour symmetric pomeron parton distributions.}

\FullConference{European Physical Society Europhysics Conference on High Energy Physics,
EPS-HEP 2009,\\
		 July 16 - 22 2009\\
		 Krakow, Poland}

\begin{document}
\newcommand{\beeq}{\begin{eqnarray}}
\newcommand{\eeeq}{\end{eqnarray}}
\newcommand{\be}{\begin{equation}}
\newcommand{\ee}{\end{equation}}
\newcommand{\bea}{\begin{array}}
\newcommand{\eea}{\end{array}}

\newcommand{\eq}{&=&}
\newcommand{\eto}{{\mbox{\textrm e}}}

\def\xp{x_{{I\!\!P}}}
\def\alfas{{\alpha_s}}
\def\pbar{\overline{p}}
\def\qbar{\overline{q}}
\def\cbar{\overline{c}}
\def\bbar{\overline{b}}
\def\dbar{\overline{d}}
\def\ubar{\overline{u}}
\def\sigmahat{\hat{\sigma}}
\def\Qbar{\overline{Q}}
\def\half{\textstyle{\frac{1}{2}}}
\def\gev{{\textrm GeV}}
\def\rbo{{\bf r}}
\def\bbo{{\bf b}}
\def\funp{{I\!\!P}}
\section{Diffractive production of $W$  bosons}

In the diffractive case, the standard inclusive parton distributions are replaced by diffractive parton 
distributions. The  asymmetry is a particularly good observable since it is
insensitive to  the gap  survival probability which in most of the approaches multiplies
both the  cross sections $d\sigma_{W^{\pm}}/dy$.
The charge asymmetry  in the $W$ boson decays at Tevatron is presented in \cite{Abe:1994rj}.
Forward-backword asymmetry of electron charged pairs is discussed in \cite{Abe:1996us}. 
Basic papers on diffractive hadroproduction of $W^{\pm}$ and $Z^0$ bosons and also dijets at high energies are: \cite{Covolan:1999sw,Covolan:2002kh}.
The single diffractive dissociation (SD), which we consider from now on, can  be interpreted as a proton-pomeron ($p\funp$) collision, where pomeron is a vacuum quantum number object with the partonic structure decribed by totaly symmetric
pomeron parton distributions
\be\label{eq:pomeronsym}
u_{\funp}(x)=\bar{u}_{\funp}(x)=d_{\funp}(x)=\bar{d}_{\funp}(x)=s_{\funp}(x)=\bar{s}_{\funp}(x)=\ldots\equiv q_\funp(x)\,.
\ee 
Thus the $W$ production cross section are related to quark distributions in the following way
\beeq\nonumber
\frac{d\sigma_{W^+}}{dy} &\sim& (u_{p}(x_1)\,+\,\dbar_{p}(x_1))\,q_\funp(x_2/\xp)
\\\label{eq:ppom}
\frac{d\sigma_{W^-}}{dy} &\sim& (d_{p}(x_1)\,+\,\ubar_{p}(x_1))\,q_\funp(x_2/\xp)\,,
\eeeq 
where the longitudinal momentum fractions, $x_1=\frac{M_W}{\sqrt{s}}\, {\rm e}^{y}$ for the proton quark and $x_2=\xp\beta=\frac{M_W}{\sqrt{s}}\, {\rm e}^{-y}$ for the pomeron quark, and
$\xp={M^2_{D}}/{s}$ is a fraction of the proton's momentum transferred into the diffractive
system of mass $M_{D}$ which contains the $W$ boson. In addition, $\beta$ is a fraction of the pomeron
momentum carried by the parton taking part in the $W$ boson production. From the condition: $0<\xp, \beta<1$, we have for the $W$ boson rapidity
\be\label{eq:gap}
-y_{max}+\ln(1/\xp)<y<y_{max}\,.
\ee
where $\Delta=\ln(1/\xp)$ is the lenght of the rapidity gap.

\begin{figure}[t]
\begin{center}
\includegraphics[width=6cm]{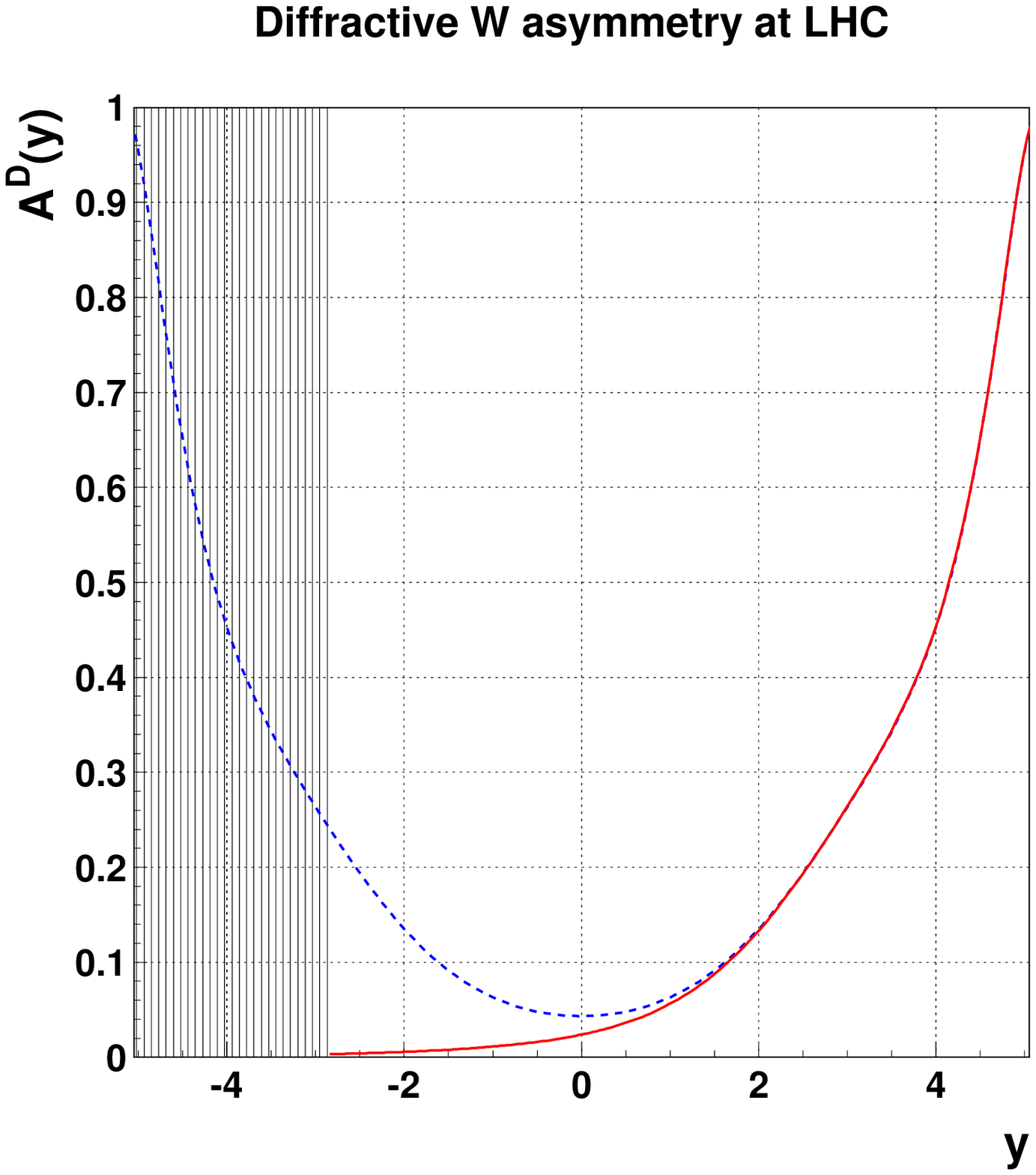}
\includegraphics[width=6cm]{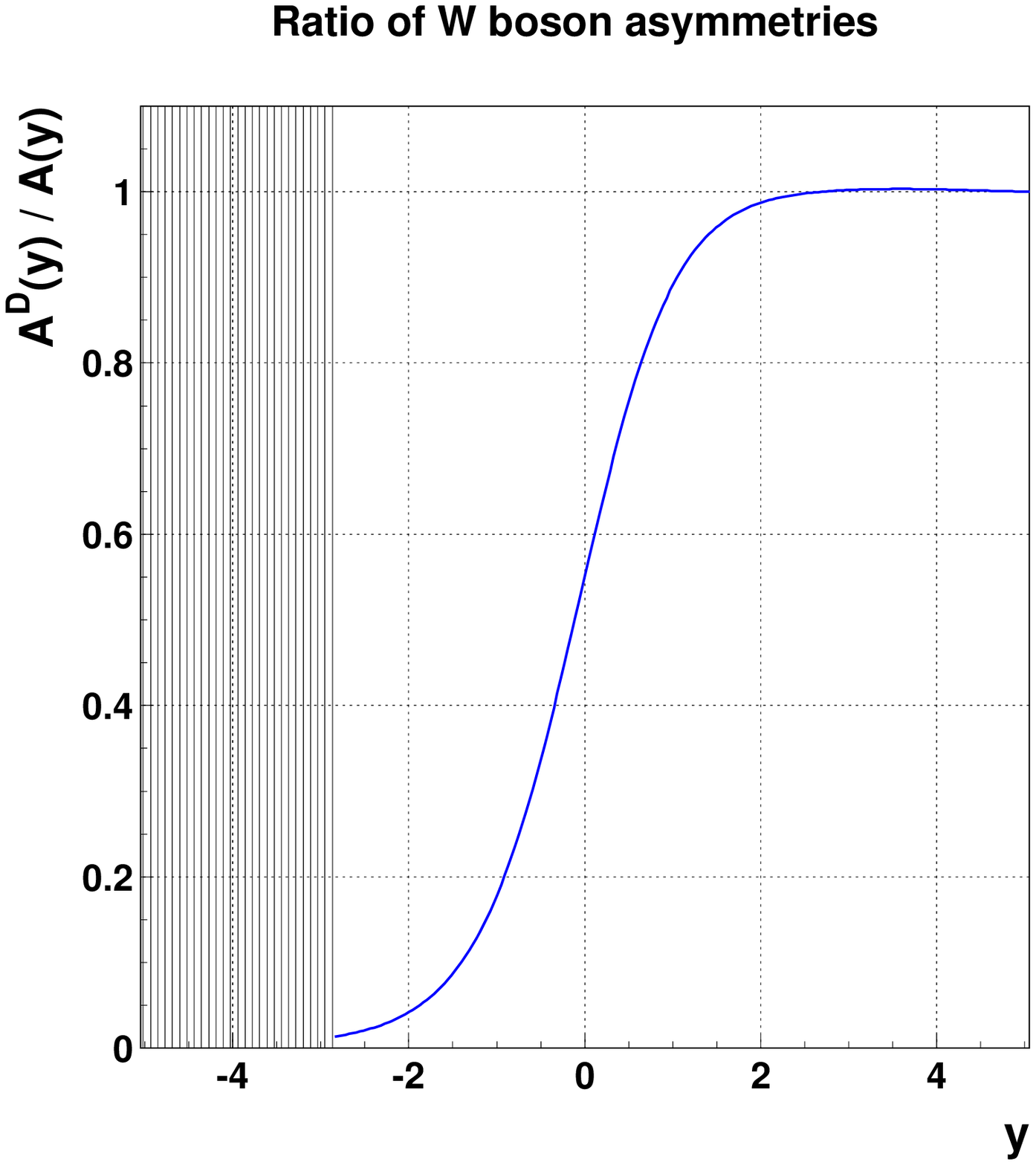}
\caption{Left: The $W$ boson asymmetry  in $p\funp$ collisions (solid line) together with the  asymmetry in  $pp$ collisions (dashed line). The shaded area on both plots indicates  the rapidity gap corresponding to $\xp=0.1$. Right: The ratio of the $W$ boson production asymmetries in the diffractive and nondiffractive $pp$ scattering.}
\label{fig:3}
\end{center}
\end{figure}

The $W$ boson production asymmetry is a particularly good observable since it is
insensitive to  the gap  survival probability \cite{Bjorken:1992er}
which  multiplies both the $W^{\pm}$  cross sections 
The pomeron parton distributions also  cancel, and we obtain
for the $W$ asymmetry in the diffractive case,
\be\label{eq:asympom1}
A^{D}(y)=\frac{u_{p}(x_1)-d_{p}(x_1)\,+\,\dbar_{p}(x_1)-\ubar_{p}(x_1)}{u_{p}(x_1)+d_{p}(x_1)
\,+\,\dbar_{p}(x_1)+\ubar_{p}(x_1)}\,,
\ee
where  the parton distributions are taken at the scale $\mu=M_W$. Notice that $A^D(y)$ is independent of
$\xp$, i.e. the length of the rapidity gap.
Substituting the decomposition written below
\beeq\nonumber
u_p(x)\eq u_{val}(x)+u_{sea}(x)\,,~~~~~~~~~~~\ubar_p(x)=u_{sea}(x)
\\\label{eq:valsea}
d_p(x)\eq d_{val}(x)+d_{sea}(x)\,,~~~~~~~~~~~\dbar_p(x)=d_{sea}(x)\,,
\eeeq
 we find diffractive asymmetry in terms of the valence and sea quark distributions 
\be\label{eq:asympom2}
A^{D}(y)=\frac{u_{val}(x_1)-d_{val}(x_1)}
              {u_{val}(x_1)+d_{val}(x_1)\,+\,2\,(u_{sea}(x_1)+d_{sea}(x_1))}\,.
\ee
This is an exact result obtained  only under the assumption (\ref{eq:pomeronsym}).
In  Fig.~\ref{fig:3} (left) we show the asymmetry (\ref{eq:asympom2}) (solid line) together  with 
the $W$ boson asymmetry in the inclusive case (dashed line), given in \cite{GolecBiernat:2009pj}. The ratio of these two asymmetries is shown on the right.

The pattern of the ratio is quite general and depends only on the assumption on flavour symmetry of the pomeron  parton distributions, Eq.~(\ref{eq:pomeronsym}). Therefore, it would be  interesting to test experimentally the very concept of the flavour  symmetric pomeron parton distributions  by measuring the ratio 
of the two $W$ asymmetries in the diffractive and nondiffractive $pp$ scattering.
If it is true, the $W$ asymmetry in the single diffractive case provides an additional constraint for the parton distribution functions in the proton.


\section{Acknowledgments}
I am indebted to Krzysztof Golec-Biernat for collaboration on the subject of this presentation.


\end{document}